# Simulation of chemical bond distributions and phase transformation in carbon chains

C.H.Wong[1], E.A. Buntov[1], V.N. Rychkov[1], M.B. Guseva[2], A.F. Zatsepin[1]

[1]*Institute of Physics and Technology, Ural Federal University, Ekaterinburg, 620002, Russia.*

[2]*Faculty of Physics, Moscow State University, Moscow, 119991 Russia.*

**Abstract:**

In the present work we develop a Monte Carlo algorithm of the carbon chains ordered into 2D hexagonal array. The chemical bond of the chained carbon is computed from 1K to 1300K. Our model confirms that the beta phase is more energetic preferable at low temperatures but the system prefers the alpha phase at high temperatures. Based on the thermal effect on the bond distributions and 3D atomic vibrations in the carbon chains, the bond softening temperature is observed at 500K. The bond softening temperature is higher in the presence of interstitial doping but it does not change with the length of nanowire. The elastic modulus of the carbon chains is 1.7TPa at 5K and the thermal expansion is +7 x $10^{-5}$ $K^{-1}$ at 300K via monitoring the collective atomic vibrations and bond distributions. Thermal fluctuation in terms of heat capacity as a function of temperatures is computed in order to study the phase transition across melting point. The heat capacity anomaly initiates around 3800K.

## 1. Introduction:

The carbon nanomaterials had been studied in the past few decades [1,2]. It provides a large variety of applications in daily life. In 1985 the discovery of $C_{60}$ molecules in the shape of a football made major contribution in medical application [1]. Another kind of carbon material, carbon nanotube, was proven to give impressive elastic modulus [2]. Carbyne seems like another strong material as well [3]. A recent theoretical study of Young's modulus of carbyne, a parallel carbon chains with kinks, gives a breakthrough at over 1TPa which draws a lot of attentions to material scientists [3]. We therefore develop a novel Monte Carlo algorithm to model 10 carbon chains in form of hexagonal array.  Thermal expansion is essential important to nano-electronics [4] and different thermal elongations of the nanomaterials make a strong impact on the reliability of the nano-electronics [2]. However, the sign of thermal expansion is different in various carbon materials [5,6,7]. The coefficient of thermal expansion of fullerene is positive [7] while the coefficient turns into negative in free standing graphene due to out-of-plane vibrations [5]. Despite the sophisticated concepts behind the thermal expansion in various carbon materials, the arrangement of different types of covalent bond is one of the most crucial parameters to determine the thermal expansion. Following the arguments of energy minimization, our work enables to study the thermal effect on the chemical bond and atomic distributions of the carbon chains and presumably identifies the bond softening temperature, thermal expansion and elastic modulus. As the variety of bond distributions alternate the energy state of the carbon chains, the average energy will be analyzed in beta and alpha phase respectively. In the second part of the simulation, we will study the factors affecting the bond softening temperature of the carbyne. The melting transition across the carbyne via heat capacity as a series of temperatures will be implemented as well. Many Monte Carlo calculations of the material science only limits in dimensionless temperature such as magnetic spin interactions [11-14,17,20,21], therefore a plenty of theoretical works have been bounded into DFT only to predict the physical properties of carbyne [8,9,10]. However, using Monte Carlo method to study the carbyne is not fully established and therefore our group creates an alternative path to make it works in the Kelvin! The simulation involves 10 carbon chains ordered in the 2D hexagonal array and each nanowire carries 50 carbon atoms (unless otherwise specify).

*Corresponding author. E-mail: ch.kh.vong@urfu.ru (Chi Ho Wong)

## 2. Simulation model:

The first part of the Monte Carlo simulation calculates the redistribution of the covalent bond at various temperatures but meanwhile readjusting the new atomic coordinates with help of Metropolis algorithm [11-14]

Assume the scattering time is temperature independent, the Hamiltonian $H$ is

$$H = e^{-T/T_{bj}} \sum_{m=1}^{M}\sum_{n=1}^{N-1} \left| E_{m,n,j} e^{-\frac{\left(r_{m,n} - r_{m,n,j}^{eq}\right)}{0.5 r_{m,n,j}^{eq}}} - E_2 \right| + e^{-T/T_{bj}} \sum_{m=1}^{M}\sum_{n=1}^{N-1} J_A\left(\cos\theta + 1\right)^2 - 4\varphi\left[\sum_{n,m} (\frac{\sigma}{r})^6 - (\frac{\sigma}{r})^{12}\right]$$

where $M$, $N$, $E_2$, $T$ are the total number of chains, the total number of carbon in each chain, double bond energy and temperature respectively. The formation in single, double and triple bond corresponds to $j = 1, 2$ and $3$ respectively. The $j$ is a stochastic variable in the simulation. The $r$ is computed in Cartesian coordinate and $r_{m,n,j}^{eq}$ is equilibrium position. For example, $r_{5,18,1}^{eq}$ refers to the equilibrium position of 18$^{th}$ atom along 5$^{th}$ chain which is connected by single bond. The temperature to break the covalent bond $T_{bj}$ is determined by $E_j = k_B T_{bj}$ where Boltzmann constant $k_B = 1.38\times10^{-23} JK^{-1}$. The Van der Waal's energy $E_{vdw}$ is the only interaction between the adjacent carbon chains with the sample length $\tau_s$. Based on the hexagonal structure, periodic boundary condition applies along XY plane so that every carbon atom interacts laterally with the three nearest neighbors via Van der Waal's force as shown in Figure 1.

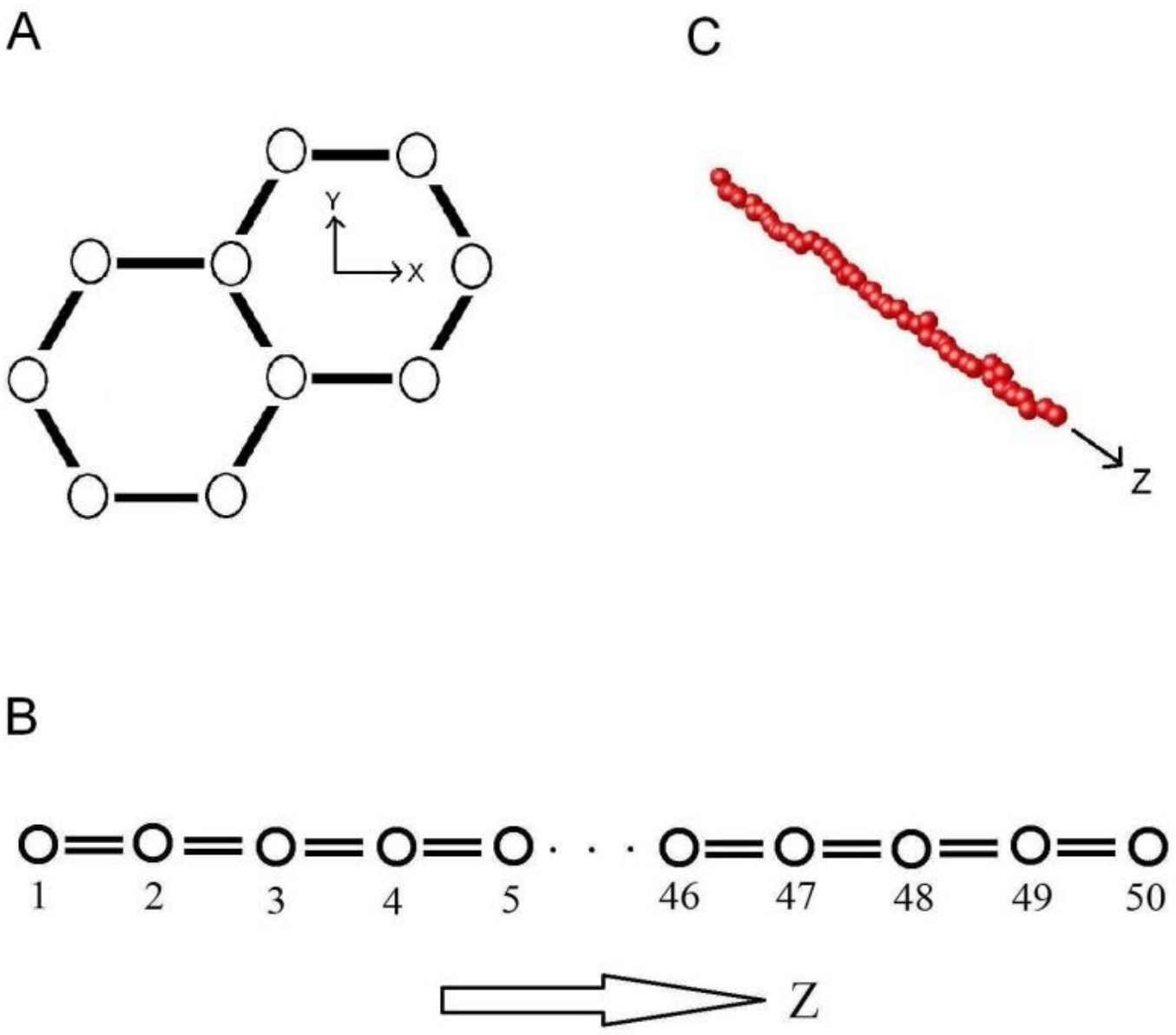

Figure 1: **A -** The cross section of the carbon chains arranged hexagonal structure. **B -** The chains are propagating along z axis at initial condition with the bond distance of 134pm (Before simulation). **C** – The simulation results of one of the carbon chains at 300K with the averaged atomic spacing of 134.8pm along Z axis (After simulation)

The initial inter-chain separation is 0.3nm. The Van der Waal's constant, $\varphi$ and $\sigma$, are calculated in combination [15,18] with $\tau_s \frac{dE_{vdw}^2}{d\tau_s^2} = \frac{1}{\zeta}$ and $\left.\frac{dE_{vdw}}{dx}\right|_{x^{eq}} = 0$. A measure of the volume changes as a response to the pressure changes at constant temperature $T$ and entropy $S$ are almost the same in solid, i.e. $-\frac{1}{V}\left(\frac{\partial V}{\partial P}\right)_T \sim -\frac{1}{V}\left(\frac{\partial V}{\partial P}\right)_S$. It makes the negligible distinction between isothermal compressibility $\zeta$ and isentropic compressibility [15] and hence $\zeta \sim \frac{1}{\rho v^2}$ is applicable here. After all, the calculation yields the $\varphi$ and $\sigma$ are $8.1\times10^{-23} J$ and $1.23\times10^{-10} m$ respectively. We set the angular energy $J_A$ to be 600kJ/mol but the actual angular energy will be weakened by $(\cos\theta+1)^2$. For example, $J_A(\cos\theta+1)^2$ equals to 0 in the straight carbon chain because the pivot angle formed by three adjacent carbon atoms is 180 degree. The single bond, double bond and triple bond energy at 300K are 348kJ/mol, 614 kJ/mol and 839 kJ/mol respectively [16]. The $C-C$, $C=C$ and $C\equiv C$ bond length are $r_{m,n,1}^{eq} = 154$ pm, $r_{m,n,2}^{eq} = 134$ pm and $r_{m,n,3}^{eq} = 120$ pm respectively [16].

In the model all carbons are initially connected by double bond and separated by 134pm. At each Monte Carlo step the carbon is selected randomly. The selected atom starts to move to new coordinate including the variation of the Van der Waal energy and also change the types of covalent bonds. The kinematics of the selected carbon with atomic mass $M$ is governed by $dz = \pm p\tau\sqrt{\frac{k_B T}{M}}$ at any temperature $T$. Here the longest scattering time $\tau$ of the atom is defined as the time to travel the covalent radius. The root-mean square-velocity of the carbon within one period of motion along Z axis is calculated by Hamiltonian of 1D harmonic-oscillator separately [15]. As a result, $\tau$ is around $1.96\times10^{-13}s$ at room temperature. As the rate of collision may amend from place to place due to stochastic process, the random number so called frictional factor $p$, varies from 0.01 to 0.99 to represents the Stochastic collision. The $dz$ is positive if the random number $R_z$ within 0 and 1 is greater than 0.5. However, once the $R_z$ is less than or equal to 0.5, the $dz$ becomes negative. Similarly, the sign of $dx$ and $dy$ are controlled by their corresponding random number $R_x$ and $R_y$ between 0 and 1 respectively. The $dx > 0$ if $R_x > 0.5$, otherwise the $dx < 0$. The same cut-off value, 0.5, is applied to the sign convention along Y axis. The out-of-chain vibration is likely weaker than the in-chain vibration. As a result, $dx = dy = \frac{k_B T}{\left(E_{m,n,1} + E_{m,n,2} + E_{m,n,3}\right)/3} dz$. Another random number $0 \le R_{bond} \le 1$ controls the types of covalent bonds. The selected $C = C$ double bond is allowed to switch into $C \equiv C$ triple bond ($R_{bond} \ge 0.5$) or $C - C$ single bond ($R_{bond} < 0.5$).

When temperature becomes higher, the selected site may be switched to single bond, triple bond or remain in double bond according to the energy minimization and Octet rule[13]. Only 8 electrons are allowed in the outermost shells in presence of the Octet rule and no lone pair electron is allowed to generate. We also include the possibility that once the double bond is excited to either single or triple bond, it can revert back to double bond. The strength of covalent bond is softened by thermal energy in parallel. If the energy difference $E_{diff}$ is less positive, the selected atom is allowed to be in motion and/or change the types of covalent bonds simultaneously. Otherwise, it returns to the previous status [17]. Thermal energy is another routine to influence the selected atom [11]. If the new random number $0 \le R_B \le 1$ is smaller than Boltzmann probability $e^{-E_{diff}/k_B T}$ [12], the selected carbon will move and/or amend the types of covalent bonds as well. The process will continue until equilibrium. As a remark, the $k_{m,n,j}$, $T_{bj}$ and $x^{eq}_{m,n,j}$ will be amended if the types of covalent bonds is swapped. The interactions are effective in the nearest neighbor only. Periodic boundary condition is applied along XY plane such that each carbon interacts with 3 nearest lateral neighbors via Van der Waal force. No metropolis step applies to the 1st carbon atom in each chain as a fixed boundary condition along Z axis. The Monte Carlo simulation is iterated 100000 times (unless specified otherwise) at each temperature.

The calculation of the elastic modulus is based on the Hamiltonian below. The work $W_{external}$ exerted on the chains is defined as the product of the mechanical deformation $dz$ and applied force $F_z$ along the chain axis.

$$H = e^{-T/T_{bj}} \sum_{m=1}^{M} \sum_{n=1}^{N-1} \left| E_{m,n,j} e^{-\frac{\left(r_{m,n} - r_{m,n,j}^{eq}\right)}{0.5 r_{m,n,j}^{eq}}} - E_2 \right| + e^{-T/T_{bj}} \sum_{m=1}^{M} \sum_{n=1}^{N-1} J_A \left(\cos\theta + 1\right)^2 - 4\varphi \left[ \sum_{n,m} \left(\frac{\sigma}{r}\right)^6 - \left(\frac{\sigma}{r}\right)^{12} \right] + W_{external}$$

Given that cross section area of the hexagonal carbyne is known, the elastic modulus is computed by comparing the chain length with and without applied force. The applied force acting on the 10 parallel chains is 10nN along Z axis. For the sake of obtaining the pure signals of the heat capacity anomaly of the α-carbyne, any thermal noise arising from the fluctuations in atomic positions is suggested to neglect in the Hamiltonian. The overall fluctuations of the heat capacity above 2000K at equilibrium state are calculated with help of $\Delta C = \frac{1}{S} \sum_{v>95000}^{MCS} \frac{\langle H^2 \rangle - \langle H \rangle^2}{k_B T^2}$ where S is the number of samplings. Applying the metropolis methodology without shifting the atomic positions away from the equilibrium coordinates, the Monte Carlo step of 100000 are iterated at each temperature. The computation of the heat capacity initiates at the steps higher than 95000.

## 3. Results

Figure 1C demonstrates the average atomic coordinate of the carbon chain. The simulated bond length along the Z axis is 134.8pm at 300K. Figure 2 show that the average normalized energies as a function of Monte Carlo steps at 100K and 900K respectively. Both curves become flat beyond 50000 steps to achieve equilibrium. By comparing the mean energies at equilibrium status, the average normalized energy at 900K is more positive than the relaxed energy at 100K. Chemical bond distributions and thermal excitations make robust impacts on the energy at equilibrium and therefore we decide to make a closer look on the bond distributions as a function of temperatures in Figure 3. The formation of double bond is dominant at low temperatures. The double bond is switched to either single bond or triple bond upon heating. However, a remarkable second downturn (upturn) of the double bond (single /triple bond) is observed at 500K. The total number of single bond exceeds the distributions of double bond above 540K. Finally the increase (decrease) of single/triple bond (double bond) turns into noticeably slower again over 600K. The model indicates that the single and triple bonds are evenly distributed at any temperature. The elastic modulus along the chain axis equals to 1.7TPa at 5K which is in the same range of some early theoretical mechanical results of the carbyne chain [3,29].

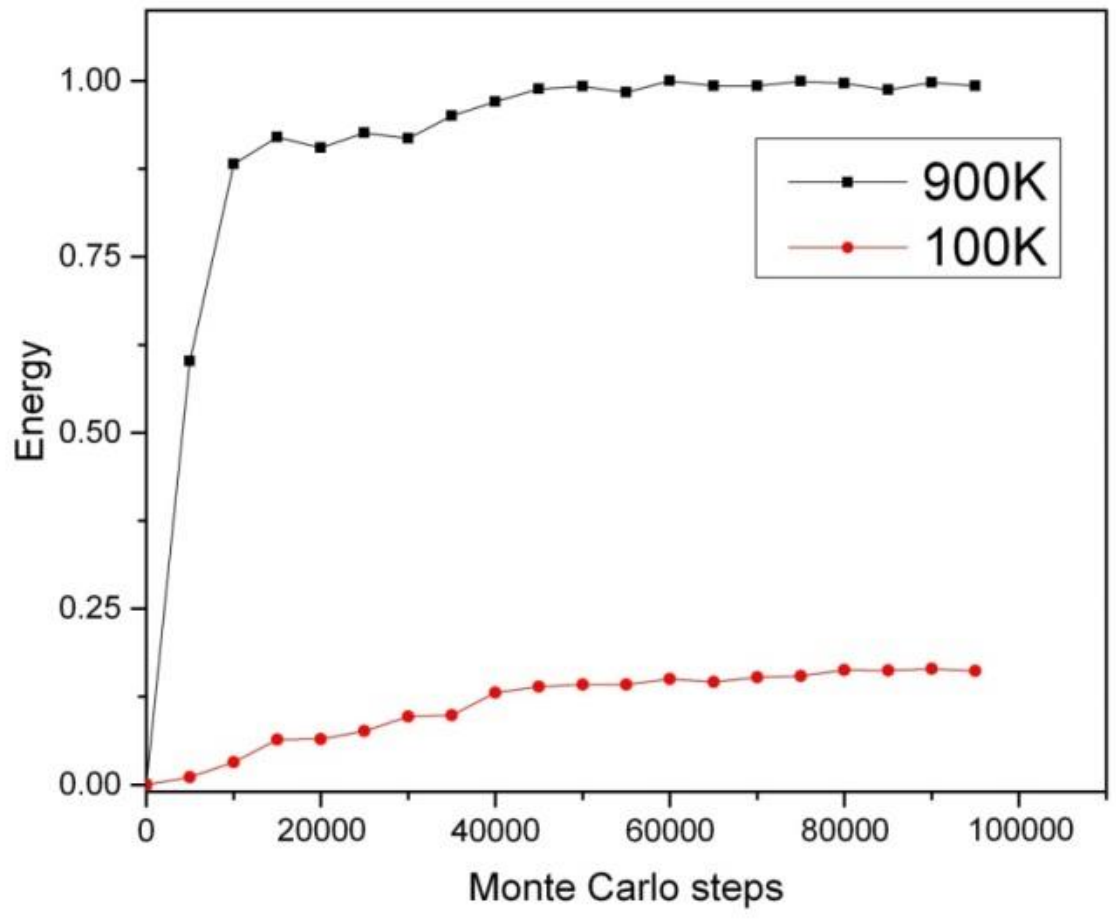

Figure 2: The average normalized energy is relaxed to equilibrium at different temperatures.

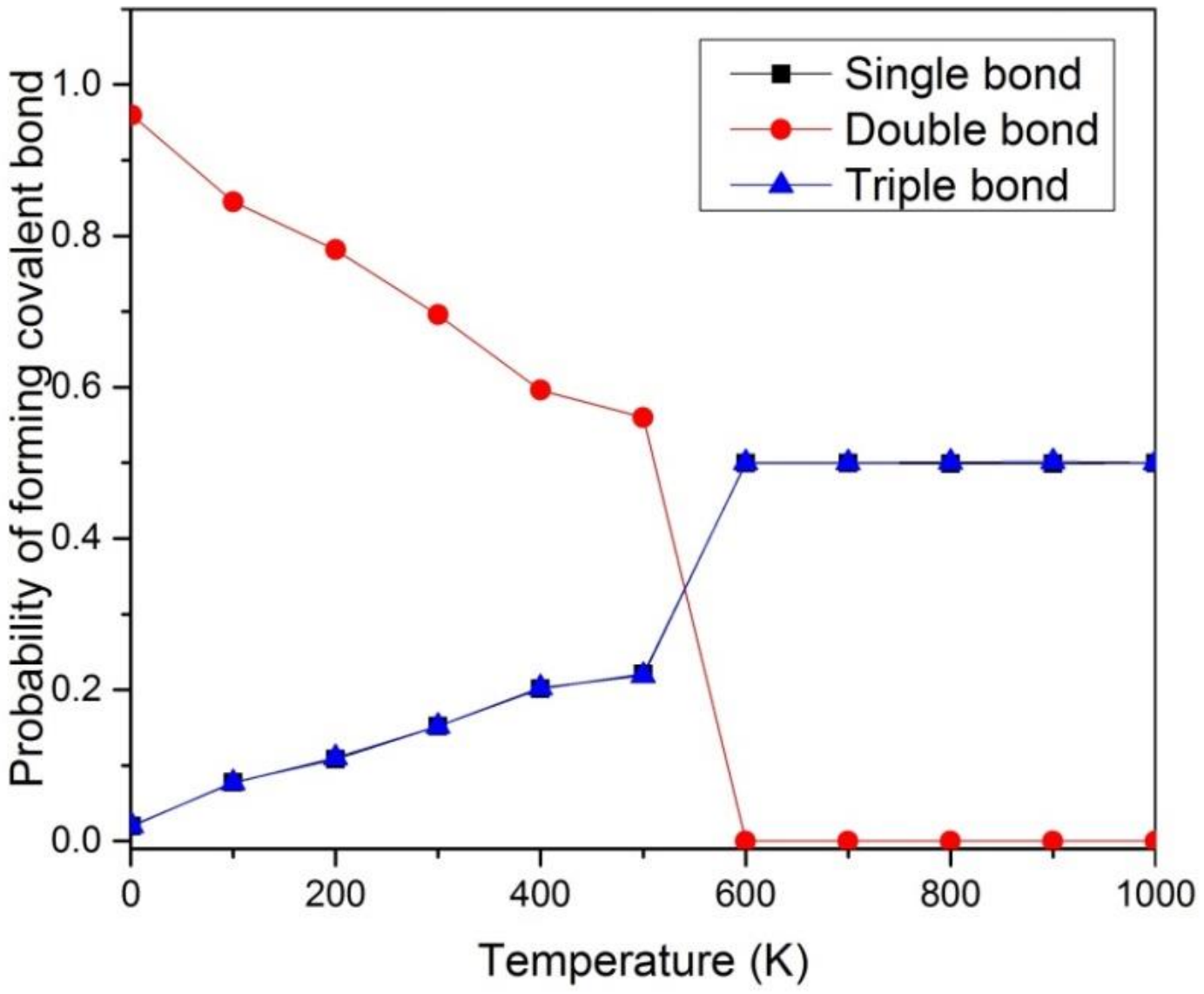

Figure 3: The probability of retaining double bonds reduces at higher temperatures. More single bonds and triple bonds are established at hotter condition.

The mean occupation of the covalent bonds at 1K:

| | | | | | | | | | | | | | | | | | | | | | | | | | | | | | | | | | | | | | | | | | | | | | | | | |
|---|---|---|---|---|---|---|---|---|---|---|---|---|---|---|---|---|---|---|---|---|---|---|---|---|---|---|---|---|---|---|---|---|---|---|---|---|---|---|---|---|---|---|---|---|---|---|---|---|
| 2 | 2 | 2 | 2 | 2 | 2 | 2 | 2 | 2 | 2 | 2 | 2 | 2 | 2 | 2 | 2 | 2 | 2 | 2 | 2 | 2 | 2 | 2 | 2 | 2 | 2 | 2 | 2 | 2 | 2 | 2 | 2 | 2 | 2 | 2 | 2 | 2 | 2 | 2 | 2 | 2 | 2 | 2 | 2 | 2 | 2 | 2 | 2 | 2 |

The mean occupation of the covalent bonds at 1000K:

| | | | | | | | | | | | | | | | | | | | | | | | | | | | | | | | | | | | | | | | | | | | | | | | | |
|---|---|---|---|---|---|---|---|---|---|---|---|---|---|---|---|---|---|---|---|---|---|---|---|---|---|---|---|---|---|---|---|---|---|---|---|---|---|---|---|---|---|---|---|---|---|---|---|---|
| 1 | 3 | 1 | 3 | 1 | 3 | 1 | 3 | 1 | 3 | 1 | 3 | 1 | 3 | 1 | 3 | 1 | 3 | 1 | 3 | 1 | 3 | 1 | 3 | 1 | 3 | 1 | 3 | 1 | 3 | 1 | 3 | 1 | 3 | 1 | 3 | 1 | 3 | 1 | 3 | 1 | 3 | 1 | 3 | 1 | 3 | 1 | 3 | 1 |

Figure 4: Scheme for chemical bond distribution of the carbon chain at different temperatures.

Figure 4 displays the chemical bond distributions along the carbyne chains. The single, double and triple bonds are abbreviated as “1”, “2” and “3” respectively. All carbon atoms are connected by $... = C = C = C = ...$ at 1K. However, when temperature is increased to 1000K, the dominant bond distribution is $... - C \equiv C - C \equiv ....$ . According to Figure 4, most carbon atoms are linked by double bonds at low temperatures but the chemical connections turn into single bond and triple bonds alternatively at high temperature. The increase of the chain length from 280K to 320K yields the coefficient of thermal expansion along the chain axis at +7 x $10^{-5}$/K with help of the central differentiation relative to 300K. Figure 5 shows the reduction in the $... = C = C = C = ...$ as a function of temperatures in different concentrations of interstitial dopants. The mean free path of the carbon atom is abbreviated as MPF where 0.71MPF, 0.54MPF and 0.49MPF correspond to the reduction of 29%, 46% and 51% in the mean free path respectively. The bond softening temperature increases from 500K to 800K when the mean free path is shortened to 0.49MPF. Figure 6 demonstrates the double bond distributions in different chain lengths. It seems like the bond softening temperatures are the same despite the nanowire is longer! One of the traditional ways to predict the melting point is to simulate the heat capacity anomaly at a series of temperatures. The fluctuations of the normalized heat capacity initiate at 3500K as shown in Figure 7. The heat capacity at each temperature is smoothed by 15 adjacent data points. As the melting point is far above 1000K and no double bond exist above 600K with the evidence of Figure 3, the simulation avoids reverting to double bond from either single or triple bond for the temperature between 2000K to 8000K in Figure 7, in order to minimize the computational cost, except for the refreshment of the double bond at initial condition at each temperature.

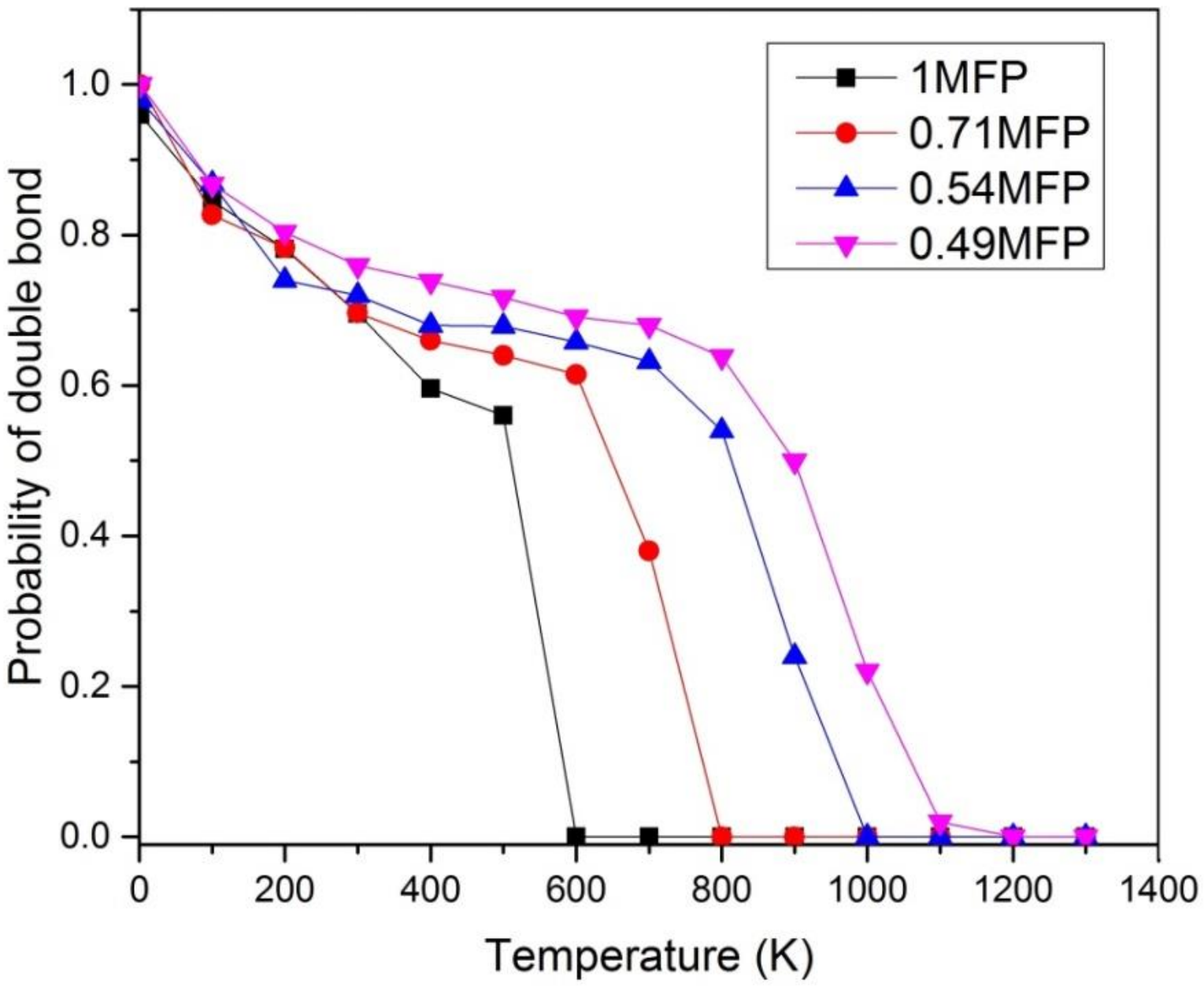

Figure 5: The double bond distributions as a series of temperatures. The bond softening temperature increases with interstitial doping. The mean free path of the atom, MPF, controls the doping levels where 0.71MPF, 0.54 and 0.49MPF refers to the reduction of 29%, 46% and 51% in the mean free path respectively.

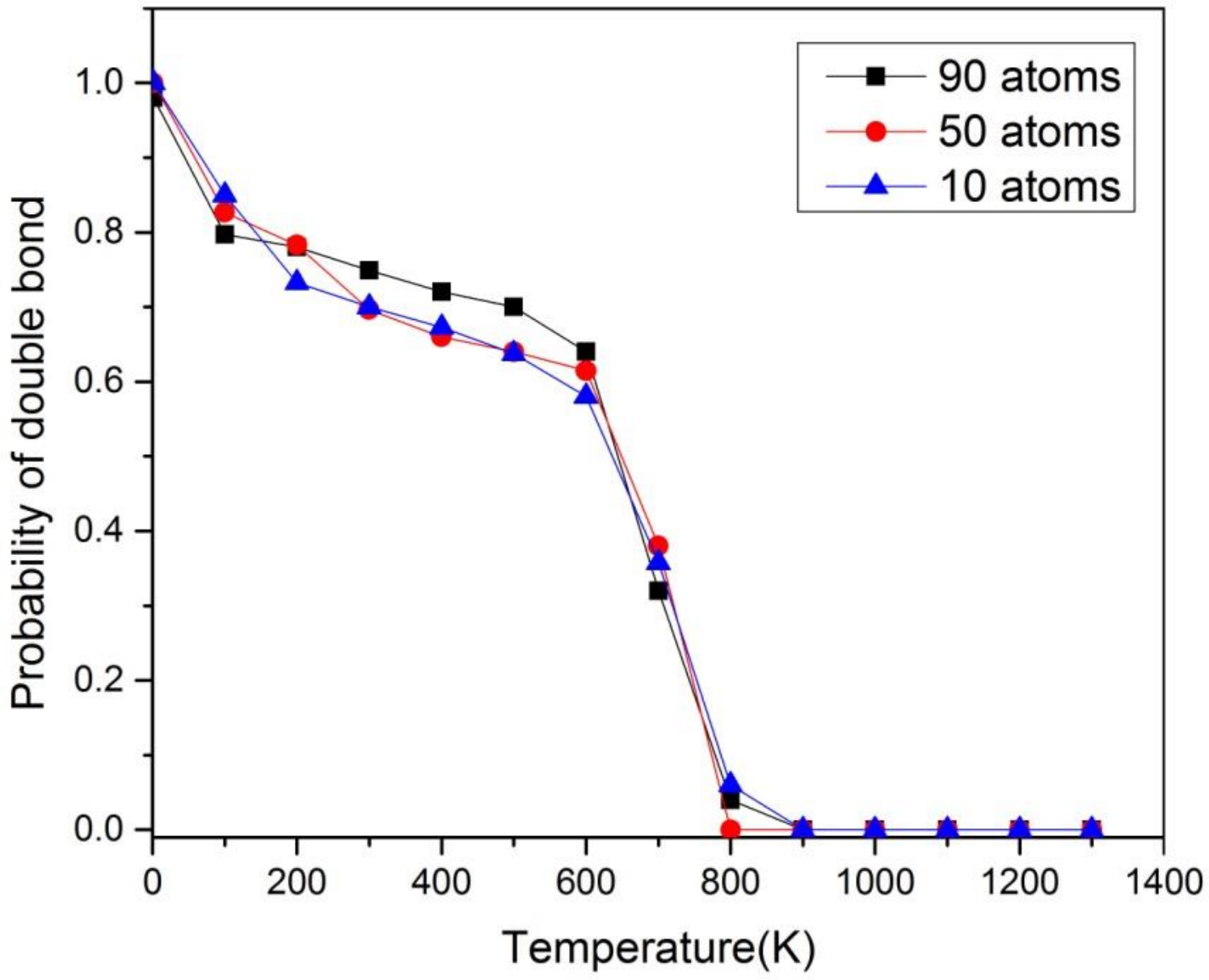

Figure 6: The probability of forming double bond as a function of temperatures in the case of 0.71MPF. The size effect on the chain length is not noticeable.

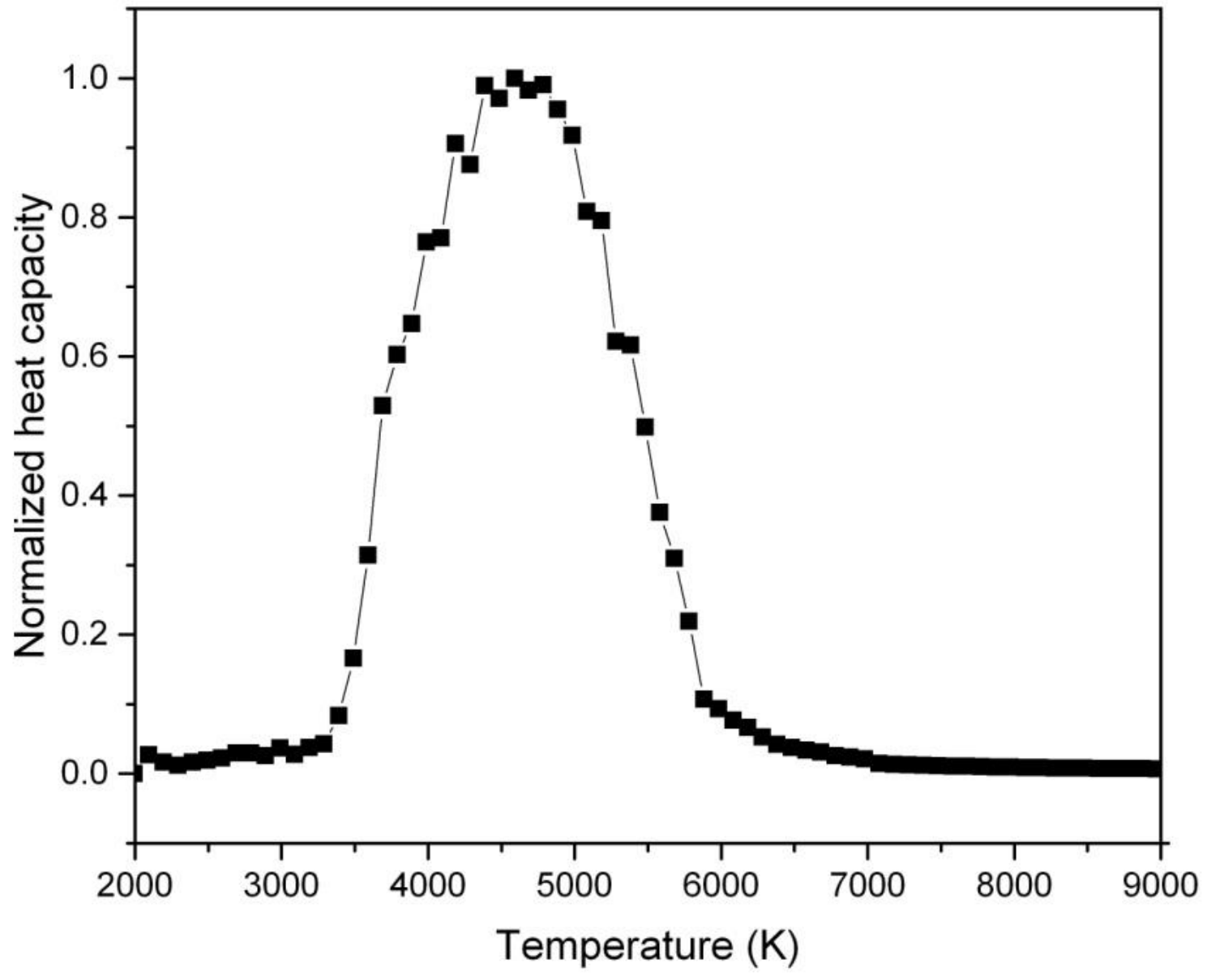

Figure 7: The normalized heat capacity anomaly as a function of temperatures. The anomaly refers to the melting transition.

**4. Discussion**

The average normalized energy at 900K as shown in Figure 2 is more positive than the energy at 100K which is owing to the more aggressive thermal excitation at high temperatures. The initial condition of the Monte Carlo simulation guides all carbons which are connected by double bond and separated by the ideal bond length of 134pm. Therefore the energy at the first Monte Carlo step is very close to zero. When the simulation starts, the atoms may change their type of covalent bonds and atomic positions which cause the energy more positive based on the Hamiltonian. However the metropolis algorithm is looking for the energy minimization simultaneously and presumably the increase of the energy starts saturation at equilibrium. The system at 900K achieves equilibrium much sooner because the Boltzmann excitation is dominant at high temperatures. Despite the probability of generating double bonds is weaker upon heating due to the Boltzmann thermal excitation, the robust reduction in double bond as shown in Figure 3 is observed at 500K. The reason is likely due to the phase transition across the β-phase to α-phase [24]. The ratio of the single to triple bond is always close to 1 because the carbon should obey Octet rule. There is another interesting feature where the decrease (increase) of double bond (single/triple bond) development turns into less aggressive again above 600K. This trendy has arisen from the saturation in the re-establishment of the chemical bonds at high temperatures. According to the simulation data, the thermal expansion at room temperature is $+7 \times 10^{-5}$/K .The order of magnitude in the coefficient of thermal expansion (CTE) is comparable to other room temperature material data [16,19] such as diamond (CTE: 1 x $10^{-6}$/K), graphite (CTE: 4 x $10^{-6}$/K) and single wall carbon nanotube (CTE: 2 x $10^{-6}$/K). Despite no reliable measurement of the elastic modulus of isolated carbyne chain is performed experimentally, both DFT [29] and our Monte Carlo simulation predict that the elastic modulus of the carbyne should be in TPa range even these two algorithms are not the same. It gives a further support to the material scientists who keep improving the fabrication process of the carbyne, in order to produce a very strong material in the world. All carbons are eventually connected by double bond at 1K as illustrated in Figure 4 because this phase minimizes the energy more effectively, with the evidence from Figure 2. The bond softening temperature turns into higher in the presence of interstitial defect as demonstrated in Figure 5. This phenomenon is credited to the shortening of the mean free path. The phase transition from beta to alpha carbyne involves the rearrangement of the bond lengths. If the mean free path of the carbon is shortened by the defect, the kinematic range of the carbon is restricted and hence more thermal energy is required to make the phase transition success. As the mean free path of the carbon should not depend on the length of the nanowires [16,18], the bond softening temperature does not change with sizes in Figure 6. The early study of the phase diagram of the carbon material gives the melting point of the carbyne at 3800K [24] and so our Monte Carlo data in Figure 7 is more or less agreed with the early report.

## 5. Conclusion

In summary, the macroscopic properties (thermal expansion and elastic modulus) of the linear carbon chains have been studied by microscopic chemical bond distributions. We evaluated these parameters using Monte Carlo Metropolis algorithm, by taking into account of both the spatial and thermal effects on the bond strength and the chaotic movement among carbon atoms. Our simulation not only shows a positive thermal expansion with coefficients of +7 x $10^{-5}$/K at room temperature, but also demonstrates the formation of single bond and triple bond is the energetic favorable at high temperatures. The bond softening temperature is unrelated to the size but it does depend on the level of interstitial dopants. In addition, our theoretical elastic modulus of the carbyne chain is 1.7TPa at 5K. This Monte Carlo algorithm guides us to tune the bond softening temperature and melting point of the carbyne. The results outlined in this paper can be treated as a milestone to manage the Monte Carlo analysis of carbon materials in the physical unit.

**Acknowledgements:**

The work was supported by Act 211 of Government of the Russian Federation, contract № 02.A03.21.0006